\documentclass[letterpaper]{ametsoc}
\usepackage{enumerate}

\journal{jas}
\usepackage{newtxtext}
\usepackage[varg]{newtxmath}
\def\smath#1{\text{\scalebox{.8}{$#1$}}}

\def\sfrac#1#2{\smath{\frac{#1}{#2}}}

\usepackage[hidelinks]{hyperref}

\bibpunct{(}{)}{;}{a}{}{,}

\title{Zonal-mean atmospheric dynamics of slowly-rotating terrestrial planets}

\authors{G. J. Colyer\correspondingauthor{Harrison Building, North Park Road, Exeter, EX4 4QF, UK}
and G. K. Vallis}

\affiliation{College of Engineering, Mathematics and Physical Sciences\\University of Exeter, UK}

\email{see College directory}

\abstract{
The zonal-mean atmospheric flow of an idealized terrestrial planet is analyzed using both numerical simulations and zonally symmetric theories,  focusing largely on the limit of low planetary rotation rate.  Two versions of a zonally symmetric theory are considered, the standard Held--Hou model, which features a discontinuous zonal wind at the edge of the Hadley cell, and a variant with continuous zonal wind but discontinuous temperature. The two models have different scalings for the boundary latitude and zonal wind. Numerical simulations are found to have smoother temperature profiles than either model, with no temperature or velocity discontinuities even in zonally symmetric simulations. Continuity is achieved because of the presence of an overturning circulation poleward of the point of maximum zonal wind, which allows the zonal velocity profile to be smoother than the original theory without the temperature discontinuities of the variant theory. Zonally symmetric simulations generally fall between the two sets of theoretical scalings, and have a faster polar zonal flow than either. Three-dimensional simulations that allow for eddy motion fall closer to the scalings of the variant model. At very low rotation rates the maximum zonal wind falls with falling planetary rotation rate, even in the three-dimensional simulations, and collapses completely at zero rotation. Nevertheless, the low-rotation limit of the overturning circulation is strong enough to drive the temperature profile close to a state of nearly constant potential temperature.
}

\begin{document}

\maketitle

\section{Introduction}

The theory of the Hadley cell has long been an object of study.  Of both historical and scientific note is the famous paper by  Hadley  on the trade winds over 250 years ago \citep{Hadley35},  and the work a  century or so later by \citet{Ferrel59} and \citet{Thomson1892}.  Ferrel introduced the notion of a second cell (now called the Ferrel cell), but none of these authors were able to give a proper explanation of the limited latitudinal extent of the Hadley cell, which they generally envisioned to extend to the pole.  Baroclinic instability was implicitly considered to be a limiting factor in the Hadley cell extent in the discussion of \citet{Lorenz67}, but even without that instability an ideal Hadley cell cannot extend to the pole. The reason for that comes from the conservation of angular momentum in the polewards-moving branch of the Hadley cell, as noted by \citet{ESchneider77}, which in the absence of frictional effects leads to the development of very strong zonal winds.   Noting that result, \cite{HH} developed a zonally symmetric theory in which, neglecting eddies and any time-dependence, they posited a circulation at low latitudes in which the total zonal specific angular momentum is conserved by the flow, and a purely zonal flow at high latitudes in thermal wind balance with the specified forcing. Matching conditions are applied at the boundary between the low- and high-latitude regions, and the satisfaction of these determines the boundary latitude $\theta_H$.   Their model may be regarded as a theory for an `ideal' axisymmetric Hadley circulation, and one of its main contributions was to show that even in the absence of baroclinic instability  the Hadley cell would not reach the pole, at least on a rapidly rotating planet like Earth. 

\cite{HH} expressed their theory in fairly general terms, but focused on the limit $\theta_H \ll 1$, which corresponds to high planetary rotation rate $\Omega$.   The low-$\Omega$ limit of the theory was then specifically considered by \citet{Hou84}, with \citet{Covey} and \citet{MV} performing a number of related simulations, looking at superrotation in particular and motivated in part by Titan. 
Various extensions to the theory relevant to Earth have also taken place.  For example, 
 \citet{LindzenHou} and \citet{PlumbHou} considered hemispherically asymmetric forcing, and \citet{CPM} extended the zonally symmetric theory, a Boussinesq model originally, to compressible atmospheres.  

In this paper, and motivated partly by Venus (which has an obliquity of only $3^\circ$), we revisit the hemispherically symmetric case, focusing on the form of the matching conditions between the low and high-latitude regions and on the low-rotation regime. In the zonally symmetric theory, we evaluate an alternative matching condition to that of \citet{HH}, which exchanges the discontinuity in zonal wind for one in temperature. We compare the theoretical predictions and obtain scalings for both theories in the low-$\Omega$ limit. In this limit, it is natural to use the boundary \emph{co}-latitude $\varphi_H\equiv{\pi}/{2}-\theta_H\ll 1$. We go on to perform both zonally symmetric and 3D simulations using an idealized GCM,  comparing the simulations with the theories and discussing how they may be reconciled, in particular by adjustments in the polar region.  To facilitate comparisons between theory and numerical simulation, we use only a simple forcing in the numerical simulations, that of \citet{HS}, which is quite similar to that of \citet{HH}.  Both use Newtonian relaxation of the temperature field towards a specified equilibrium that is maximum at the equator and minimum at the poles, without diurnal or seasonal variation.

An outline of the paper follows. Section \ref{theory} contains a summary of the Held--Hou theory, including discussion of the matching conditions and the general solution. Section \ref{next_after_HH} introduces the alternative matching condition and its consequences. The two theories are compared further in section \ref{comparison}, and then in section \ref{low-rot} the low-$\Omega$ limit is considered and various scalings obtained. Section \ref{numerical} describes the numerical modeling and results, with comparison to theory and consequent discussion. In section \ref{venus} we draw attention to some implications for the idealized modeling of Venus, but we leave the actual study of Venus to a later paper. We conclude in section \ref{conclusions}. An Appendix provides more detail concerning the conversion used to compare the theory with the simulations.

\section{The Held--Hou theory for zonally symmetric atmospheres}
\label{theory}

\subsection{Summary of derivation}

We first summarize the theory of \citet{HH} (hereafter HH). Readers who are familiar with it may wish to skip to section \ref{next_after_HH}, referring back to this section as needed.
HH start from the Boussinesq version of the hydrostatic primitive equations on a 
sphere. (The compressible hydrostatic primitive equations in pressure co-ordinates have the same form as the Boussinesq equations \citep{AOFD2}, so the Boussinesq approximation is not as restrictive as it may seem.)
Steady flow  ($\partial/\partial t = 0$, where $t$ is time) and zonal symmetry ($\partial/\partial\phi = 0$, where $\phi$ is longitude) are assumed throughout, and we will only use the equations in their inviscid approximation. The two horizontal components of the momentum equation are
\begin{align}
\frac{Du}{Dt} - fv - \frac{uv\tan\theta}{a} & = 0 
\label{DuDt}\\
\frac{Dv}{Dt} + fu +  \frac{u^2\tan\theta}{a} & = - \frac{1}{a}\frac{\partial\Phi}{\partial\theta},
\label{DvDt}
\end{align}
where $\theta$ is latitude, $z$ is height above the surface, $a$ is the planetary radius, $\boldsymbol{v}(\theta,z) = (u,v,w)$ is the flow velocity, $f = 2\Omega\sin\theta$ is the Coriolis parameter, $\Phi$ is the geopotential and
$D /Dt = (v/a) \partial / \partial\theta + w \partial/ \partial z$.  
The equations are completed by the incompressibility condition
\begin{equation} 
\nabla\cdot\boldsymbol{v} = \frac{1}{a\cos\theta}\frac{\partial (v\cos\theta)}{\partial\theta} + \frac{\partial w}{\partial z} = 0,
\label{incompressibility}
\end{equation}
 hydrostasy
\begin{equation}
\frac{\partial\Phi}{\partial z}= \frac{g\Theta}{\Theta_0},
\end{equation}
where $g$ is the acceleration due to gravity and $\Theta$ is potential temperature, and
 the thermodynamic equation
\begin{equation}
\frac{D\Theta}{Dt}= -\frac{(\Theta-\Theta_E)}{\tau}.
\label{DThetaDt}
\end{equation}
Radiative-convective equilibration is represented by the thermal forcing term on 
the right-hand side that relaxes the potential temperature towards a specified equilibrium $\Theta_E$ (see (\ref{HHzavgforcing}) below) with a specified time $\tau$.

At the top of the atmosphere $z = H$ we take $w = 0$, and (\ref{DuDt}) may then be written
\begin{equation}
v(\zeta+f) = 0,
\label{dichotomy}
\end{equation}
where
\begin{equation}
\zeta = -\frac{1}{a}\frac{\partial u}{\partial\theta} + \frac{u\tan\theta}{a}
\label{zmrelvor}
\end{equation}
is (the vertical component of) the relative vorticity. Equation (\ref{dichotomy}) 
may be satisfied in two ways: (i) $v = 0$ (identically), which by (\ref{incompressibility}) implies $w = 0$,  self-consistently; we will call  $\boldsymbol{v} = (u,0,0)$ a circulation-free solution; or (ii) total vorticity $\zeta+f = 0$, equivalent to $\partial M/\partial\theta = 0$, conserving the total zonal specific angular momentum $M\equiv (u + \Omega a\cos\theta)a\cos\theta$; this has solution
\begin{equation}
u = u_M(\theta) + \frac{u(0)}{\cos\theta},
\label{momentum-conserving-solution}
\end{equation}
where
\begin{equation}
u_M \equiv \frac{\Omega a\sin^2\theta}{\cos\theta}.
\label{uM}
\end{equation}
The alternatives (i) and (ii) will be applied to high- and low-latitude
regions respectively (note that (\ref{uM}) would diverge at the pole), with matching conditions at the boundary latitude $\theta_H$, which is the edge of the Hadley cell. (In some monsoon contexts the alternatives may both apply at low latitudes but in different regimes, as in \citealt{Geen}.)

The remaining dynamical equations may be combined by cross-differentiating---assuming that the advective term $\boldsymbol{v}\cdot\nabla v$ in (\ref{DvDt}), which is zero in case (i), is also small 
in case (ii)---to produce an equation for thermal gradient wind balance, and integrating in $z$---assuming that $u$ is small in the boundary layer---to get at $z = H$
\begin{equation}
fu + \frac{u^2\tan\theta}{a} = -\frac{gH}{a\Theta_0}\frac{\partial\bar\Theta}{\partial\theta},
\label{TWtop}
\end{equation}
where the overbar denotes the vertical average ${H^{-1}}\int_0^H dz$. The first and second terms on the left-hand side correspond to the geostrophic and cyclostrophic thermal wind gradients respectively.

In the circulation-free case (i), we now use the thermodynamic equation (\ref{DThetaDt}) to conclude that $\Theta = \Theta_E$. Equation (\ref{TWtop}) must then be solved for $u\equiv u_E$ given $\bar\Theta = \bar\Theta_E$, which is specified as
\begin{equation}
\frac{\bar\Theta_E}{\Theta_0} =1 - \sfrac{2}{3}\Delta_H P_2(\sin\theta),
\label{HHzavgforcing}
\end{equation}
where $P_2(x) = (3x^2 - 1)/2$ and $\Theta_0$ and $\Delta_H$ are parameters. The solution is \begin{equation}
u_E = \Omega a\cos\theta\left(\sqrt{2R + 1} - 1\right),
\label{uE}
\end{equation}
where $R\equiv \Delta_H g H/\Omega^2 a^2$, as in HH. We note that this $u_E$ corresponds to global rigid-body rotation about the planetary axis, i.e. constant angular velocity $u_E/a\cos\theta = \sqrt{2R + 1} - 1$. This is a consequence of the specified form of \eqref{HHzavgforcing}.

In the angular-momentum-conserving case (ii), $u$ is given by (\ref{momentum-conserving-solution}), and (\ref{TWtop}) becomes an equation to be solved for $\bar\Theta$.
The solution is
\begin{equation}
\frac{\bar\Theta(0) - \bar\Theta}{\Theta_0} = \frac{u^2 - u(0)^2}{2gH}.
\label{TWsol}
\end{equation}
This immediately implies a finite upper bound on $u$, 
since $\bar\Theta > 0$. Specializing to the case $u(0)=0$,\footnote{ 
This eliminates both equatorial superrotation, $u(0) > 0$, 
in accordance with \citet{Hide69},
and non-equatorial extrema in $\bar\Theta$, which occur where $u=0$ if $u(0) < 0$.
}
(\ref{TWsol}) becomes
\begin{equation}
\frac{\bar\Theta(0) - \bar\Theta}{\Theta_0} = \frac{u_M^2}{2gH} = \frac{\Omega^2 a^2\sin^4\theta}{2gH\cos^2\theta}, 
\label{TWsolspec}
\end{equation}
which is equation (12) of HH.

\subsection{Matching conditions}

HH proposed the following two matching conditions: continuity of temperature,
\begin{equation}
\bar\Theta(\theta_H-) = \bar\Theta_E(\theta_H+),
\label{continuity}
\end{equation}
and the closure of the energy budget over the Hadley cell, integrating (\ref{DThetaDt}) and using $\boldsymbol{v}\cdot\nabla\Theta=\nabla\cdot(\boldsymbol{v}\Theta)$, implied by (\ref{incompressibility}), to obtain
\begin{equation}
\int_0^{\theta_H}\bar\Theta\cos\theta d\theta= \int_0^{\theta_H}\bar\Theta_E\cos\theta d\theta.
\label{closure_orig}
\end{equation}
Since $\Theta = \Theta_E$ in the high-latitude region, this region may be added to the integral to write 
\begin{equation}
\int_0^{{\pi}/{2}}(\bar\Theta-\bar\Theta_E)\cos\theta d\theta = 0.
\label{closure_full}
\end{equation}
The second matching condition implies $\bar\Theta\sim\bar\Theta_E\sim\Theta_0$ 
and hence that the upper bound on $u\sim\sqrt{gH}$.

\begin{figure}[t]
\centering
\includegraphics[trim={0 20 0 20},clip,scale=1.0]{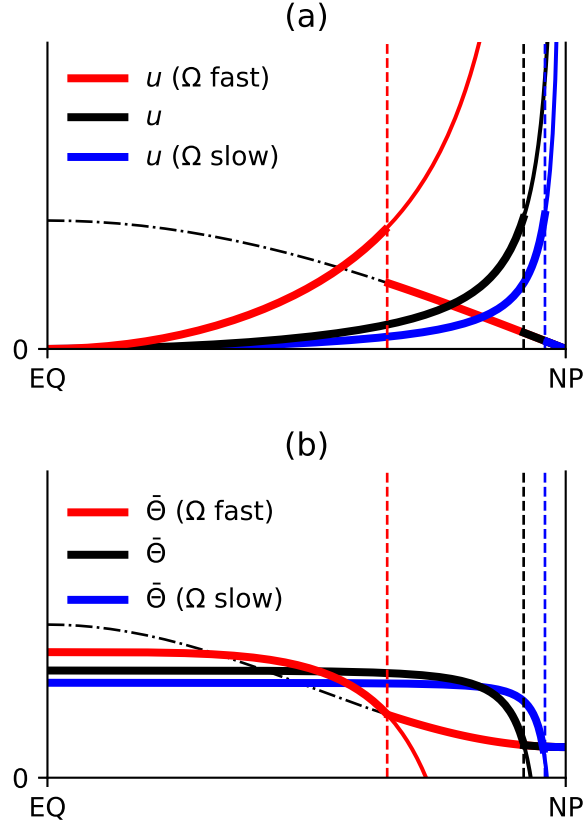} 
\caption{Sketch of the Held--Hou model (cf. Figs. 1 and 3 of HH) for three different planetary rotation rates $\Omega$, indicated by different colours. (a) Zonal wind versus latitude: the solid line shows the angular-momentum-conserving zonal wind $u_M$, which diverges at the pole; the dot-dashed line shows the zonal wind $u_E$ that is in thermal wind balance with $\bar\Theta_E$;
(b) potential temperature: the solid line shows the temperature $\bar\Theta$ in thermal wind balance with
$u_M$; this too diverges at the pole; the dot-dashed line shows the forcing temperature $\bar\Theta_E$. The original HH matching conditions determine the latitude (vertical dashed line in both panels) at which one solution crosses over to the other, where $\bar\Theta$ is continuous but $u$ is discontinuous (thick solid lines).
}
\label{sketches_HH}
\end{figure}

\subsection{General solution} 

HH focused mainly on the low-$R$ limit, and we will focus mainly on the
high-$R$ limit considered by \citet{Hou84}. However, we first describe the
general solution. The zonal wind $u$ is already given on either side of $\theta_H$ by (\ref{uM}) and (\ref{uE}), but $\theta_H$ is still to be determined. $\bar\Theta$ is given on either side of $\theta_H$ by (\ref{HHzavgforcing}) and (\ref{TWsolspec}), in which the only other unknown is $\bar\Theta(0)$.
These may be substituted into the second matching condition \eqref{closure_orig} which upon integration gives
\begin{equation}
\frac{\bar\Theta(0)}{\Theta_0} = 1 + \Delta_H\left[ 
\sfrac{1}{3} -\sfrac{1}{3}x_H^2\left(1+\frac{1}{2R}\right) - \frac{1}{2R}
 + \frac{1}{4Rx_H}\ln\left(\frac{1+x_H}{1-x_H}\right)
\right],
\label{offset}
\end{equation}
where $x_H\equiv\sin\theta_H$. 
The first matching condition may be used to eliminate $\bar\Theta(0)$ and obtain the expression
\begin{equation}
R = \sfrac{3}{4}\left[ 
\sfrac{1}{3} + \frac{1}{x_H^2} + \frac{x_H^2}{1-x_H^2} 
- \frac{1}{2x_H^3}\ln\left(\frac{1+x_H}{1-x_H}\right)
\right],
\label{R_HH}
\end{equation}
which is implicit in equation (17) of HH and may be inverted numerically, i.e. solved for $x_H$ given $R$. $\bar\Theta(0)$ and hence $\bar\Theta$ generally is then determined.

Fig. \ref{sketches_HH} shows a sketch of the solutions for (a) $u$ and (b) $\bar\Theta$, and how they vary with $\Omega$. The dot-dashed lines show $u_E$ and $\bar\Theta_E$ respectively. For each $\Omega$ (each colour), the solid lines show the angular-momentum-conserving solutions (which both diverge at the pole), the thick lines are the combined HH solutions, and the vertical dashed lines mark $\theta_H$---where the temperature curves cross, by the first matching condition. The second matching condition sets the equatorial intercept of temperature such that the areas under the $\bar\Theta$ and $\bar\Theta_E$ curves are equal, or equivalently that the net area between the curves is zero, when weighted by $\cos\theta$. (Note that the weighting factor is not incorporated into the sketched curves, however. It has the same functional form as $u_E$.)  A notable feature of the HH theory is that $u$ is discontinuous at $\theta_H$. This corresponds, via thermal wind balance, to the discontinuous gradient in $\bar\Theta$, even though $\bar\Theta$ itself is continuous.

As $\Omega$ decreases (red then black then blue), both $u$ and $\bar\Theta$ flatten in the Hadley-cell region, by (\ref{uM}) and (\ref{TWsolspec}) respectively, and $\theta_H$ moves poleward. To preserve the equal-area constraint, the low-latitude temperature is reduced. Although $u_M$ is also reduced at any given $\theta$, the discontinuity in $u$ at $\theta_H$ is increased. This fact, in part, motivates us to suggest an alternative matching condition, which we consider next. In section \ref{low-rot} we will develop the low-$\Omega$ scalings of the two theories quantitatively.

\begin{figure}[t]
\centering
\includegraphics[trim={0 20 0 20},clip,scale=1.0]{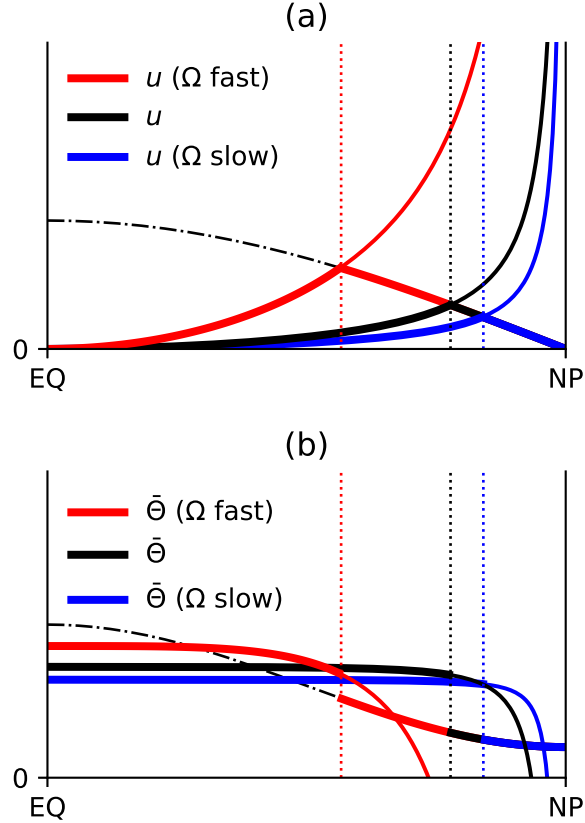} 
\caption{Sketch of the model with modified matching conditions that determine the latitude (vertical dotted line in both panels) at which one solution crosses over to the other, where $u$ is continuous but $\bar\Theta$ is not (thick solid lines).
}
\label{sketches_contu}
\end{figure}

\section{An alternative matching condition}
\label{next_after_HH}

We will see below that the discontinuity in $u$ at $\theta_H$, shown as the red curve in Fig. \ref{theory-scalings}(b), is of the same order as $u$ itself (on the equatorward side of higher $u$---and also, in the low-rotation limit, is much larger than $u$ on the poleward side). 
We could avoid this discontinuity  by changing the first matching condition to specify that $u$ itself be continuous at the boundary latitude:
\begin{equation}
u_M(\theta_H-)=u_E(\theta_H+).
\label{newcont}
\end{equation}
Using (\ref{uM}) and (\ref{uE}), this gives the boundary latitude directly:
\begin{equation}
\cos^2\theta_H = 1 - x_H^2 = \frac{1}{\sqrt{2R + 1}}.
\label{newcrossover}
\end{equation}
These two equations replace (\ref{continuity}) and (\ref{R_HH}). Equations (\ref{closure_orig}) and (\ref{closure_full}) are essentially a consequence of the dynamics, and have not been changed; (\ref{offset}) was deliberately written in a form that is independent of the first matching condition, so $\bar\Theta(0)$ is determined as before. The consequence, unsurprisingly, is that there is now a discontinuity in $\bar\Theta$ instead of $u$. We will return to discuss this trade-off later.

Fig. \ref{sketches_contu} shows a sketch similar to Fig. \ref{sketches_HH} but for the new
matching condition. The underlying $u_M$ and $u_E$ curves are all as before, but now the boundary latitude is set by where they---and not the temperature curves---cross. The boundary still moves poleward as $\Omega$ is decreased, but for all values of $\Omega$ the boundary is equatorward of the original boundary. The temperature discontinuity at the new boundary is evident. It is given by adding (\ref{HHzavgforcing}) to (\ref{TWsolspec}), using \eqref{offset}, and \eqref{newcrossover} to eliminate $R$:
\begin{equation}
\frac{\bar\Theta(x_H-)-\bar\Theta_E(x_H+)}{\Theta_0} = \frac{\Delta_H}{(2-x_H^2)}\left[ 
\frac{5}{3} - \frac{1}{x_H^2} + \frac{(1-x_H^2)^2}{2x_H^3}\ln\left(\frac{1+x_H}{1-x_H}\right)
\right].
\label{disc_contu}
\end{equation}
Again the discontinuity becomes larger as $\Omega$ is reduced. The temperature \emph{gradients} on either side of the discontinuity are equal, by thermal wind balance. In order still to satisfy the second matching condition with the new boundary, the equatorial temperature is shifted slightly higher (relative to Fig. \ref{sketches_HH}, for a given $\Omega$), to maintain the equal areas. However, this shift is always small (and actually vanishes in both the low- and high-rotation limits), as seen in Fig. \ref{theory-scalings}(e) below.

\section{Comparison of the two variants}
\label{comparison}

With the new matching condition, the ratio of the change in the gradient of $u$ at the boundary  is
\begin{equation}
\left.\frac{u_M^\prime}{u_E^\prime}\right|_{x_H} = -\left(\frac{\sqrt{2R+1}+1}{\sqrt{2R+1}-1}\right),
\label{u_gradient_ratio}
\end{equation}
which has the limit $-1$ as $R\rightarrow\infty$, and diverges $\sim-2/R$ as $R\rightarrow 0$.
With the original matching condition, the latter divergence result is the same, but both $u_M^\prime/u_E^\prime$ and $u_M/u_E$ (at the boundary) diverge $\sim-2\sqrt{2R}/3$ as $R\rightarrow\infty$.

\begin{figure}[tb] 
\centering
\includegraphics[trim={10 15 10 40},clip,scale=0.85]{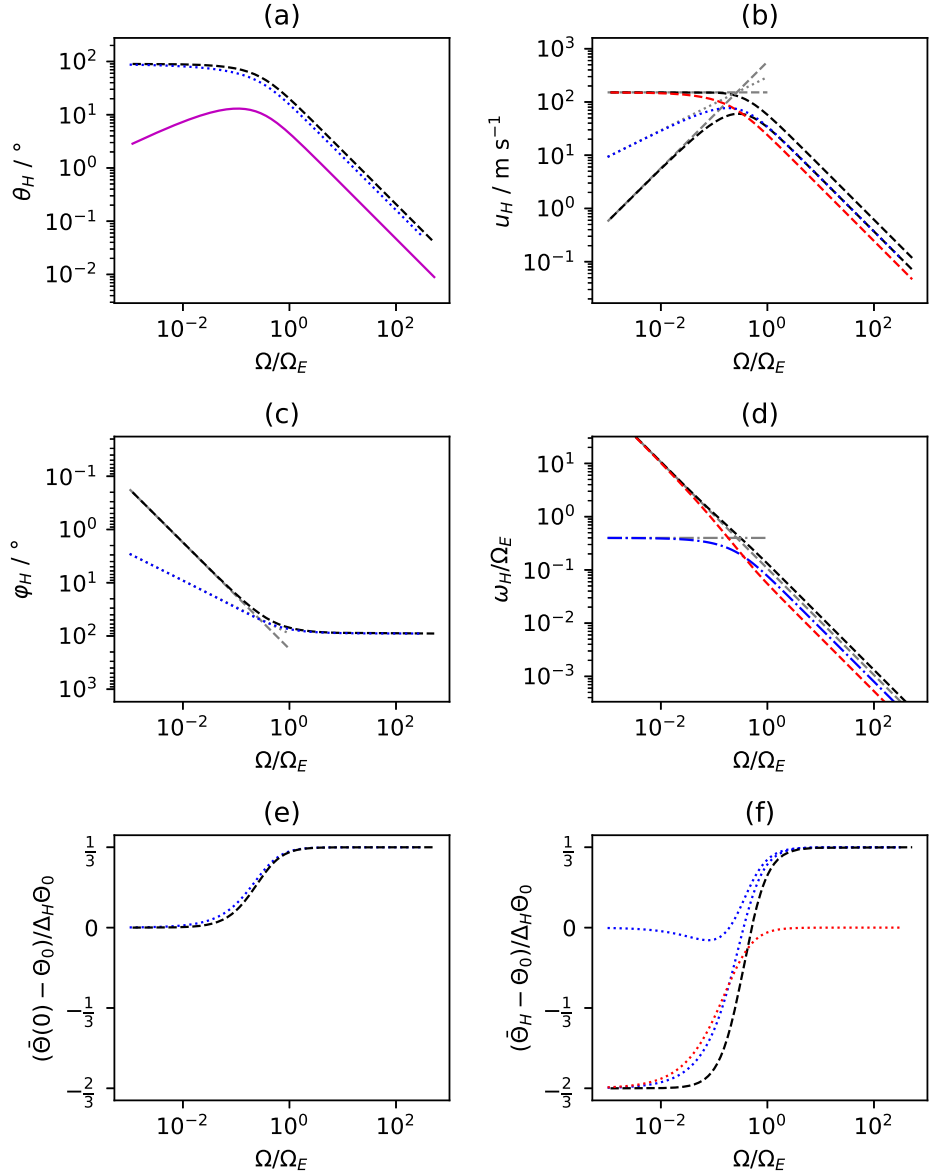} 
\caption{Theoretical solutions of the HH and continuous-$u$ theories. Parameters
appropriate for Earth have been used, so that these plots match Fig. \ref{scalings}, and the variable planetary rotation rate $\Omega$ has been normalized to Earth's, $\Omega_E$. In all panels, dashed lines correspond to the HH theory, dotted lines to the continuous-$u$ theory, and dot-dashed lines to both.
(a) Hadley cell boundary latitude, $\theta_H$; 
(b) boundary zonal wind, $u_H$; 
(c) boundary co-latitude, $\varphi_H\equiv{\pi}/{2}-\theta_H$;
(d) boundary zonal wind as an angular velocity around the planetary axis, $\omega_H\equiv u_H/a\cos\theta_H$;
(e) equatorial temperature, $\bar\Theta(0)$;
(f) boundary temperature, $\bar\Theta_H$.
The red curves in the right-hand column of panels show the discontinuity (difference) at the boundary, between the appropriate pair. The solid magenta curve in the first panel shows the difference between the two theories. Gray straight lines indicate the low-$\Omega$ asymptotes derived in the text and also shown in Fig. \ref{scalings}.
}
\label{theory-scalings}
\end{figure}

Figure \ref{theory-scalings} shows further quantitative comparison of various aspects of the two theories. Note that at high $\Omega$ (low $R$), which is the limit originally considered by HH, all of the power-law scalings shown are the same for both theories; they differ only by an order-unity prefactor. Furthermore, the boundary latitude $\theta_H$ is close across the whole range of $\Omega$ (differing by no more than $\sim 10^\circ$), as is the equatorial temperature $\bar\Theta(0)$. However, different scalings are obtained from the two theories at low $\Omega$ (high $R$), including for the boundary \emph{co}-latitude $\varphi_H$. The alternative boundary condition is therefore mainly of interest in this limit, in which it is natural to use the co-latitude $\varphi\equiv{\pi}/{2}-\theta$ instead of $\theta$. In the following section we derive the low-$\Omega$ scalings indicated by the gray lines in 
Fig. \ref{theory-scalings},  and then in section \ref{numerical} we compare them with numerical simulation results from a GCM code.

\section{The low-rotation limit and co-latitude scalings}
\label{low-rot}

\subsection{General}

We now turn to the limit $\Omega\rightarrow 0$.  In the Hadley cell we see immediately from (\ref{TWsolspec}) that, except near the poles,
\begin{equation*}
\frac{\bar\Theta(0) - \bar\Theta}{\Theta_0} = \frac{\Omega^2 a^2\sin^4\theta}{2gH\cos^2\theta}\rightarrow 0,
\end{equation*}
and so $\bar\Theta \rightarrow \bar\Theta(0) =$ const. This result was pointed out by \citet{Hou84}.
As mentioned above, it means that the temperature is flattened (relative to the forcing temperature) within the Hadley cell. This flattening is illustrated in Figs. \ref{sketches_HH} and \ref{sketches_contu}.
Towards the (north) pole it is natural to make the small-angle approximation for the co-latitude $\varphi \equiv{\pi}/{2} - \theta$, whence
\begin{equation}
\frac{\bar\Theta(0) - \bar\Theta}{\Theta_0} \sim 
\frac{\Omega^2 a^2}{2gH\varphi^2}.
\label{theta_smallcolat}
\end{equation}
For the circulation-free solution at the pole itself, (\ref{HHzavgforcing}) gives
\begin{equation}
\frac{\bar\Theta_E}{\Theta_0} =1 - \frac{2}{3}\Delta_H + \Delta_H\varphi^2 + O(\varphi^4),
\label{polar-equilibrium}
\end{equation}
which again is constant to leading order, 
and 
the second matching condition
becomes simply
\begin{equation}
\bar\Theta(0) = \Theta_0.
\label{zerozero}
\end{equation}

From (\ref{uM}), the zonal wind towards the poleward edge of the Hadley cell is
\begin{equation}
u_M(\varphi) \sim \frac{\Omega a}{\varphi}.
\label{uMcolat}
\end{equation}

On the poleward side of the boundary, we have cyclostrophic balance; taking the large-$R$ limit of (\ref{uE}),
\begin{equation}
u_E(\varphi) = \sqrt{2\Delta_H gH}\varphi.
\label{uEcolat}
\end{equation}
That $u_E\propto\varphi$ in this limit means the polar vortex rotates like a rigid body with angular velocity
\begin{equation}
\frac{u_E}{a\varphi} = \frac{\sqrt{2\Delta_H gH}}{a},
\label{polarvortex}
\end{equation}
which is independent of $\Omega$ (at fixed $\Delta_H gH$---not at fixed $R$). We noted the rigid-body rotation for this global $\bar\Theta_E$ earlier. In the present limit it depends only on the vanishing $\varphi$ and non-vanishing $\varphi^2$ terms in the polar expansion \eqref{polar-equilibrium}.
The polar vorticity (both relative and absolute) is $
2\sqrt{2\Delta_H gH}/a$ in this limit.

\subsection{Low-rotation solutions with original HH matching conditions}

We impose (\ref{continuity}) to determine the transition co-latitude $\varphi_H \equiv {\pi}/{2} - \theta_H$. Using (\ref{theta_smallcolat}), (\ref{polar-equilibrium}) and (\ref{zerozero}) we get,
to leading order,
\begin{equation}
1-\frac{\Omega^2 a^2}{2gH\varphi_H^2} = 1 - \frac{2}{3}\Delta_H,
\label{theta_smallcolat_limit}
\end{equation}
and hence
\begin{equation}
\varphi_H = \frac{\sqrt{3}\Omega a}{2 \sqrt{\Delta_H gH}} = \sqrt{\frac{3}{4R}}.
\label{transition}
\end{equation}
This last result was also given implicitly by \citet{Hou84} as part of his equation (34).

From (\ref{uMcolat}), the maximum zonal wind, at the poleward edge of the Hadley cell, is therefore
\begin{equation}
u_M(\varphi_H) = \frac{\Omega a}{\varphi_H} = \frac{2 \sqrt{\Delta_H gH}}{\sqrt{3}},
\label{uMedge}
\end{equation}
which like the polar vorticity is independent of $\Omega$, as noted by \cite{Covey}, and is consistent with 
the form of the upper bound on $u$ mentioned earlier.
At the edge of the polar vortex, from (\ref{uEcolat}),
\begin{equation}
u_E(\varphi_H) = \sqrt{2\Delta_H gH}\varphi_H = \sqrt{\frac{3}{2}}\Omega a.
\label{uEedge}
\end{equation}
Thus, the Hadley cell extends with constant $\bar\Theta$ almost to the pole; at its edge, the maximum zonal wind is independent of $\Omega$, and beyond there is a polar vortex with rigid-body rotation whose angular velocity is independent of $\Omega$.

The dimensionless parameter $R$ is similar to a thermal Rossby number, at least in the limit $R\ll 1$, which corresponds to high $\Omega$ (and to the geostrophic limit of equation (\ref{TWtop}),  in which only the first term appears on the left-hand side). In this limit, $R\sim u_E/\Omega a$, and in the Hadley cell region $R  \sim u_M/\Omega a$. 

In the opposite limit here (the cyclostrophic limit), $R\gg 1$, and $\sqrt{R}\sim u_M/\Omega a$, 
whereas in the polar region 
$u_E/\Omega a$ is of order unity; in fact, using (\ref{uMedge}) and (\ref{uEedge}), in the cyclostrophic limit
$u_M(\varphi_H)/u_E(\varphi_H) = 2\sqrt{2R}/3\gg 1$.
This is the large discontinuity in $u$ already discussed: of the same order as $u_M$, and in the low-$\Omega$ limit large compared to $u_E$ and $\Omega a$. Fig. \ref{sketches_HH} is necessarily drawn for finite $\Omega$ but so as to be suggestive of this limit.

\subsection{Low-rotation solutions with continuous $u$}

In the low-$\Omega$ limit (\ref{newcrossover}) gives
\begin{equation}
\varphi_H = \left(\frac{1}{2R}\right)^{\frac{1}{4}} = \frac{\sqrt{\Omega a}}{(2\Delta_H gH)^{\frac{1}{4}}},
\label{newtransition}
\end{equation}
and hence, using (\ref{uMcolat}) or (\ref{uEcolat}),
\begin{equation}
u(\varphi_H) = \sqrt{\Omega a}( 2\Delta_H gH )^{\frac{1}{4}}.
\label{newuedge}
\end{equation}
The $\varphi_H$ given by  (\ref{newtransition}), although larger than that given by (\ref{transition}), is still small; both approach the pole asymptotically as $\Omega\rightarrow 0$. In this limit, the closure integral (\ref{closure_full}) is
unaffected by the tiny polar region.
We will return to this point when we discuss the numerical results below.

Note that the continuous-$u$ behaviour circumvents a possible paradox in the original 
theory---namely that the maximum $u$, given by (\ref{uMedge}), stays constant even as $\Omega\rightarrow 0$. In the continuous-$u$ theory, the maximum $u$ is given by (\ref{newuedge}), and, although the polar vorticity stays constant, because the polar vortex shrinks, its boundary zonal velocity decreases to zero. One might intuitively expect to get $u = 0$ for a stationary planet, although the temperature forcing still produces a zonally symmetric and circulation-generating global structure even without rotation: the planetary axis is no longer an axis of rotation, but the poles and the equator are still extrema of the forcing in this model.

\section{Numerical modelling}
\label{numerical}

\subsection{Simulation setup}
\label{simulations}

We performed simulations using \textsc{Isca} \citep{Isca}, a modeling framework that uses the Flexible Modeling System 
of the Geophysical Fluid Dynamics Laboratory (GFDL), in Princeton.   We integrate the  hydrostatic primitive equations using a spectral dynamical core, first presenting results at T42 horizontal resolution.\footnote{For our zonally symmetric simulations (only), the T for `triangular' is strictly not appropriate, because only the $m = 0$ modes are present. However, rather than define a new and unfamiliar notation for our different spectral grids, we prefer to specify the corresponding 3D resolution first, and then state any further truncation to a reduced range of zonal modes.} The parameter values used (e.g. for atmospheric mass, gas constants etc.) are those of Earth, except for rotation rate $\Omega$ where noted.

Thermal forcing and Rayleigh damping terms are as specified by \cite{HS}, with additional $\nabla^8$ hyperviscosity at grid scales.
Several variants of the HH forcing have been used in idealized studies. In particular, static stability against small-scale convection may be introduced via a latitude-dependent term, as by \cite{HS}, or via a modified adiabatic index in the exponent relating temperature to potential temperature, as by \cite{MV}. Such functional forms would replace the $z$-dependent term in equation (2) of HH: but although this does determine the magnitude of the overturning circulation, it has played no role in our scaling analysis for the boundary latitude and zonal wind; it disappeared when vertically averaged to obtain our equation (\ref{HHzavgforcing}). In principle, more general forms would modify the above theory, but in practice the effect is small. In the simulations that we report here, we treat Held--Suarez forcing as effectively the same as HH. 
For more details, please see the Appendix.

\begin{figure}[t]
\centering
\includegraphics[trim={25 20 10 30},clip,scale=0.75]{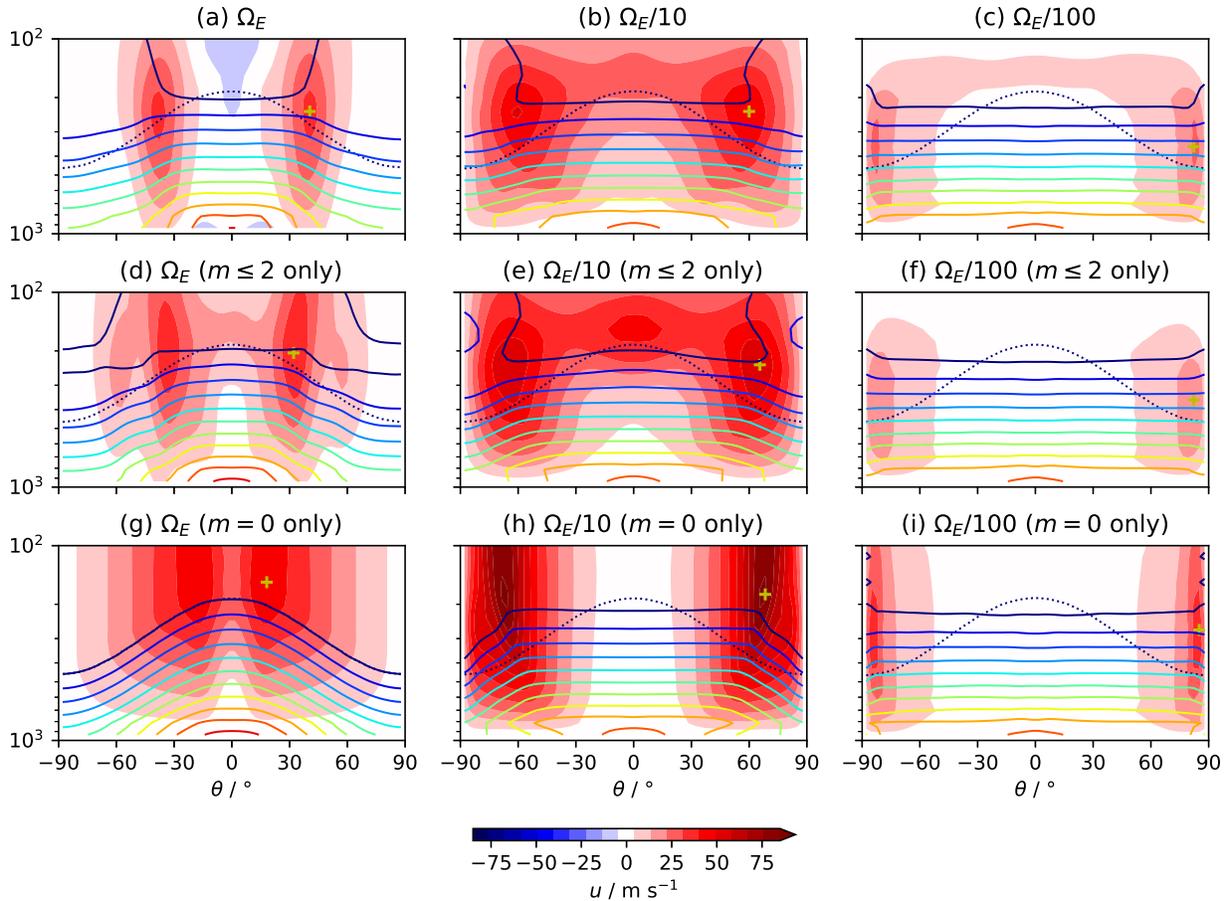} 
\caption{350-day zonal means against
latitude $\theta$ and model  
level pressure $\sigma p_{\mathrm{ref}}$ / hPa, where $p_{\mathrm{ref}} = 10^3$ hPa: zonal wind $u$, in m s$^{-1}$ (shaded colors between contour levels); temperature $T$ (solid contour lines, at 205 K and above in 10 K intervals); forcing temperature $T_{\mathrm{eq}}$ (dotted contour line, at 205 K only); and the point of maximum $u$ within the northern hemisphere (NH) and the plotted pressure range (yellow `+' sign): 
(a)-(c) full 3D simulations at T42 resolution, labelled by planetary rotation rate $\Omega$;
(d)-(f) reduced simulations truncated to zonal wavenumbers $m = \{0, 1, 2\}$ only, for the same $\Omega$ values;
(g)-(i) zonally symmetric simulations truncated to zonal wavenumber $m = 0$ only, for the same $\Omega$ values.
All simulations use 30 unevenly spaced $\sigma\equiv p/p_s$ 
levels; only the bottom decade 
is shown. 
$\Omega_E = 7.3\times 10^{-5}$ s$^{-1}$ is the rotation
rate of Earth.
}
\label{uplot}
\end{figure}

\begin{figure}[t]
\centering
\includegraphics[trim={25 20 10 30},clip,scale=0.75]{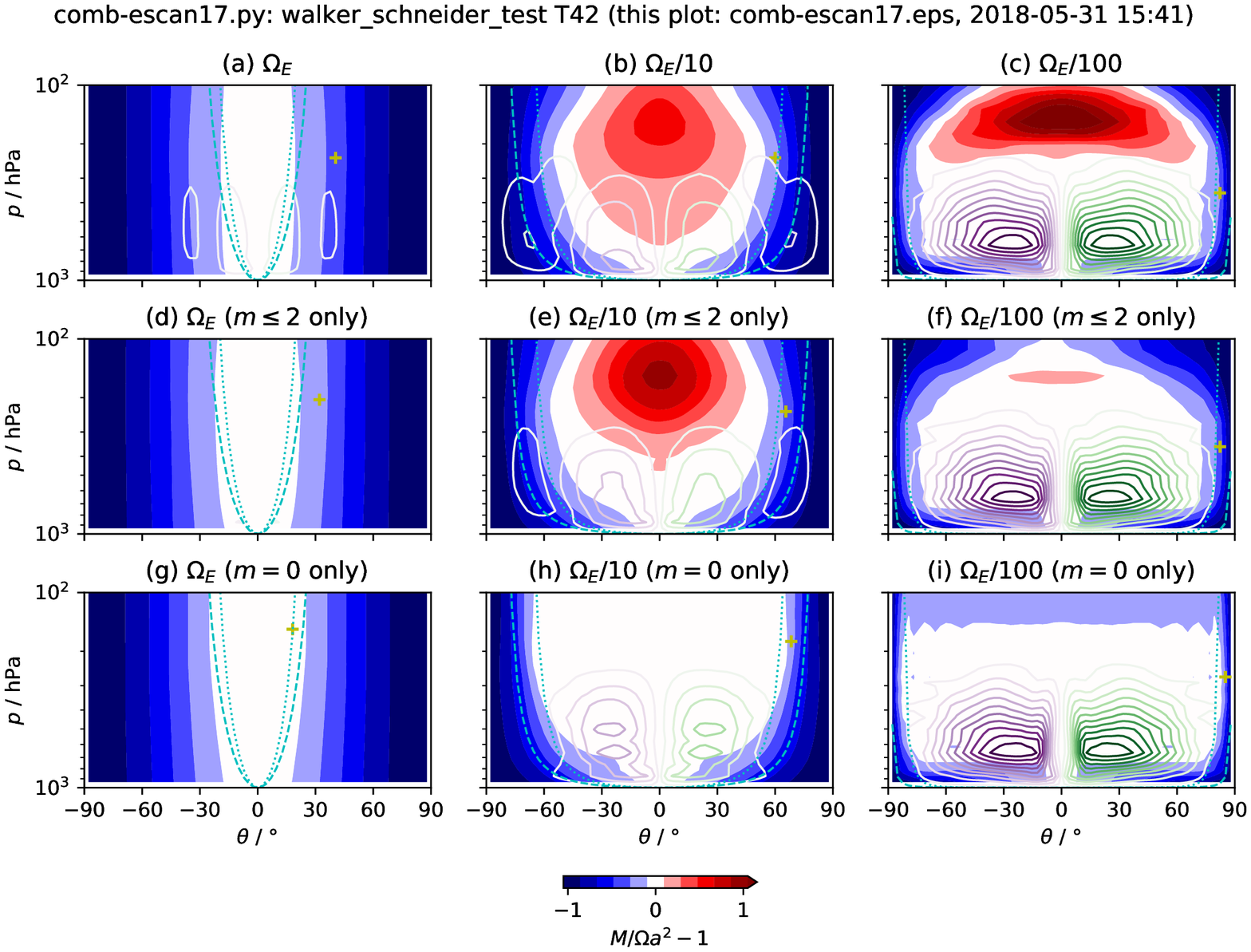} 
\caption{As Fig. \ref{uplot}, 
with $u$ converted to the total zonal specific angular momentum $M= (u + \Omega a\cos\theta)a\cos\theta$ normalized to the
equatorial planetary specific angular momentum $\Omega a^2$, where $a = 6.4\times 10^6$ m is the radius of Earth; also showing the mass streamfunction 
of the overturning circulation (contour lines, at $1.05 \times 10^{11}$ kg s$^{-1}$ intervals: equivalently, 20 levels between $\pm 1 \times 10^{12}$ kg s$^{-1}$ inclusive). The cyan curves show the predicted Hadley cell height as a function of boundary latitude, as explained in the Appendix---dashed and dotted corresponding respectively to the theory with HH matching condition and to the variant with continuous $u$.
}
\label{mplot}
\end{figure}

\subsection{Results}
\label{results}

Fig. \ref{uplot} shows simulations at three different rotation rates: Earth's $\Omega_E$ (left column), $\Omega_E/10$ (centre column) and $\Omega_E/100$ (right column). Each panel shows the zonal wind (as color shading) and the temperature (as solid contour lines). The top-left panel (a) is the Earth-like case, and each step to the right corresponds to a factor 10 reduction in $\Omega$. The rows correspond to fully 3D simulations (top row), reduced simulations truncated to zonal wavenumbers $m = \{0, 1, 2\}$ only (middle row, henceforth denoted W2), and zonally symmetric (ZS) simulations truncated to zonal wavenumber $m = 0$ only (bottom row). These last are the simulations closest to the Held--Hou model.

The physical picture underlying the Held--Hou model is made more manifest in Fig. \ref{mplot}. This shows the same zonal wind data as Fig. \ref{uplot}, but $u$ has been converted into the total zonal specific angular momentum $M= (u + \Omega a\cos\theta)a\cos\theta$, and plotted as $M/\Omega a^2 - 1$, where $\Omega a^2$ is the planetary specific angular momentum at the equator. This figure also shows contours of the overturning streamfunction.

Several trends are evident in these two figures. First, consistent with
the above theory, the Hadley-cell circulation pattern widens (and strengthens) as the planetary rotation reduces, eventually spanning almost the entire planet; the temperature flattens, and (consistent with thermal wind balance) the mid-latitude zonal jets also move poleward. These observations are true not only for the zonally symmetric simulations but for all three rows.
The white region in each panel of Fig. \ref{mplot} is the region within which the zonal angular momentum is close to its equatorial surface value, i.e. has been approximately conserved by the circulation. The red regions show superrotation. Notwithstanding the differences between the panels in this second figure, it is striking how simple the zonal wind appears when shown as $M$, compared with the complex structure of $u$ in the first figure. The additional structure in $u$ is purely geometric, as the expression for $M(u)$ makes clear.

Second, as the planetary rotation rate decreases, the 
zonal wind at first increases and then decreases, not only in the mid-latitude jets but also at low latitudes for the 3D simulations, transitioning from retrograde equatorial winds in case (a) to prograde winds in case (b), which features a wide superrotating layer. This latter case is representative of Titan, in which context such behaviour has been studied by \cite{MV} and later workers, who found that the superrotation is driven by nonlinear interactions between eddies (that is, modes with nonzero zonal wavenumber $m$). Consistent with this, superrotation is absent in the corresponding zonally symmetric simulation (h). However, below the superrotating layer, the similarity between the 3D and the zonally symmetric simulations appears to become closer as the rotation rate reduces. Although there is no superrotation aloft in case (i), it looks otherwise very similar to case (c).

The simulations that include zonal modes $m = \{0, 1, 2\}$ only, shown on the middle row of each figure, are even closer to the full 3D simulations, and in particular are able to produce superrotating flow.  We use such reduced simulations to access higher resolutions than would be computationally feasible for fully 3D simulations, as discussed further in subsection \ref{testing}.   The most obvious discrepancy between the theoretical model and these results---and again this applies to all three rows---is that the temperature does not reduce sharply to the forcing temperature at the edge of the Hadley cell. In each panel of Fig. \ref{uplot}, the lowest temperature contour shown is the 205 K contour, and the 205 K forcing contour is shown dotted in the same color. They only overlay at all in case (g), in which they overlay everywhere because the overturning circulation is so weak.  

\begin{figure}[t]
\centering
\includegraphics[trim={10 15 10 30},clip,scale=0.7]{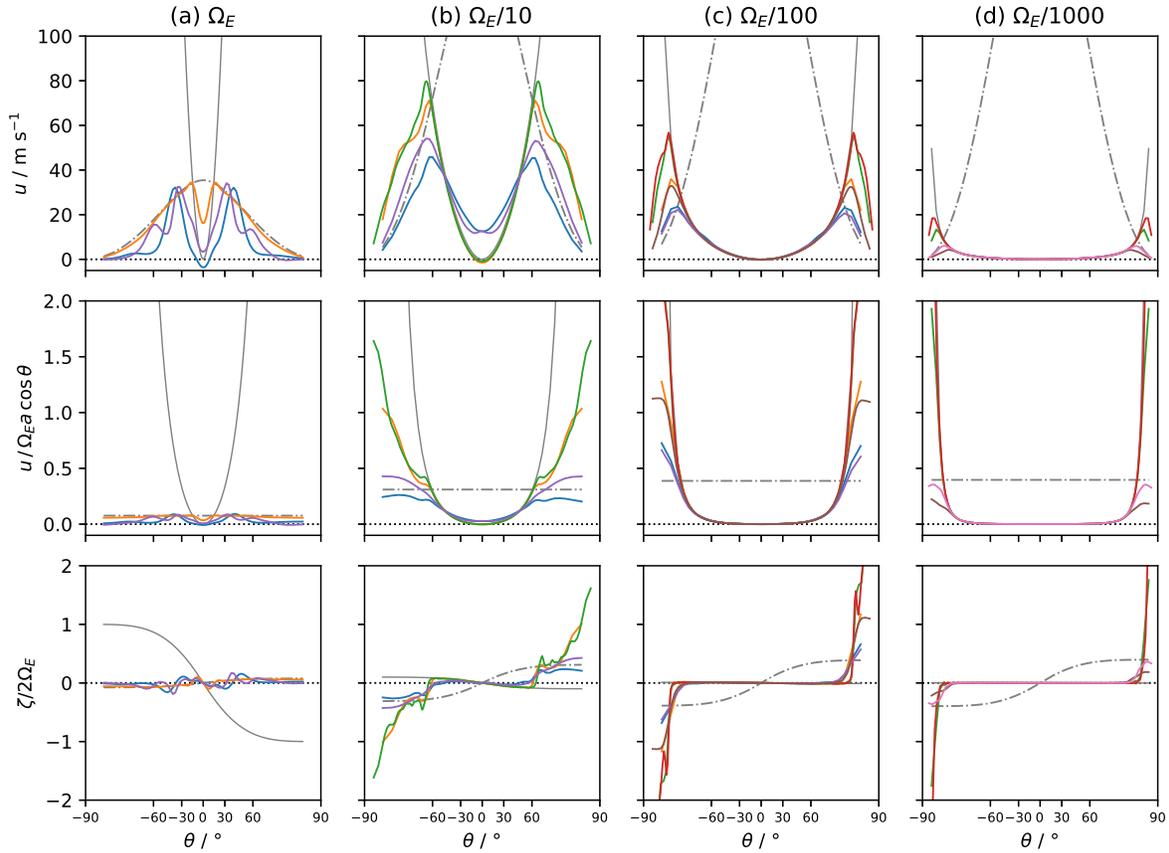} 
\caption{Horizontal profiles of 350-day zonal means of zonal velocity $u$ (top row), zonal velocity expressed as an angular velocity around the planetary axis (middle row), and relative vorticity $\zeta$ (bottom row), at model pressure level $\sigma p_{\mathrm{ref}} = 311$ hPa. Columns are labelled by planetary rotation rate $\Omega$. 
Colors match the corresponding data sets shown in Fig. \ref{scalings}. Thin gray curves indicate respectively the angular-momentum-conserving solution (solid) and the solution in thermal wind balance with the forcing temperature (dot-dashed), as explained in the Appendix. In order to expand the polar regions smoothly, the horizontal axis is nonlinear in $\theta$ but linear in $\sqrt{90^\circ-|\theta|}$ (in each hemisphere).
}
\label{hprofiles}
\end{figure}

It is also clear in the simulation results that the zonal wind does not suddenly jump to a smaller value at the Hadley cell edge.  Fig.\ \ref{hprofiles} shows horizontal profiles of the zonal wind $u$ (top row) at a single selected model level near the top of the overturning circulation. The zonal velocity $u$ does track $u_M$ well in the Hadley cell region at the lowest rotation rates (rightmost two columns) ($u_M$ is marked by the solid gray line), but decreases from its maximum quite smoothly. There is a sharper change in its gradient $\partial u/\partial\theta$, but the gradient on either side is of similar magnitude (unlike in Fig. \ref{sketches_HH} at the dashed vertical lines; see the earlier discussion in section \ref{comparison}). 
So it appears that both the zonal wind varies more smoothly than in the original theory, and the temperature varies more smoothly than in either theory.

The new matching condition may appear to be even more unphysical than the first, because a discontinuity in $\bar\Theta$ is problematic for thermal wind balance, even with equal gradients $\bar\Theta^\prime$ on either side. We might attempt to fix this in the theoretical model by adding a constant to $\bar\Theta$ in the polar region to close the discontinuity, which does not change the zonal wind (since thermal wind balance involves only the gradient), but such a fix does mean that $\bar\Theta = \bar\Theta_E$ no longer holds there. Hence, from (\ref{DThetaDt}), the polar region is no longer circulation-free in steady-state. The closure condition (\ref{closure_full}) will now include a contribution from the polar region; however, as mentioned at the end of section \ref{low-rot}, in the low-$\Omega$ limit its contribution will be negligible, both because the boundary latitude is close to $\theta = {\pi}/{2}$ and because the weighting factor $\cos\theta$ is small. In other words, a tiny adjustment to the Hadley cell covering the rest of the planet would bring the system back into global energy balance.

More generally, \emph{any} adjustment to the temperature would make a negligible contribution to (\ref{closure_full}) in the low-$\Omega$ limit. The adjustment would not, however, reproduce the same $u_E$, and hence would not in general leave the boundary latitude or the polar vorticity unaffected. Since a generalized model of this form is under-determined in the absence of any additional principles, a more comprehensive understanding of the polar region is needed.  Before addressing this issue further in subsection \ref{reconciling}, we compare the simulation results with the theoretical low-$\Omega$ scalings more quantitatively.

\begin{figure}[t]
\centering
\includegraphics[trim={10 25 10 55},clip,scale=0.85]{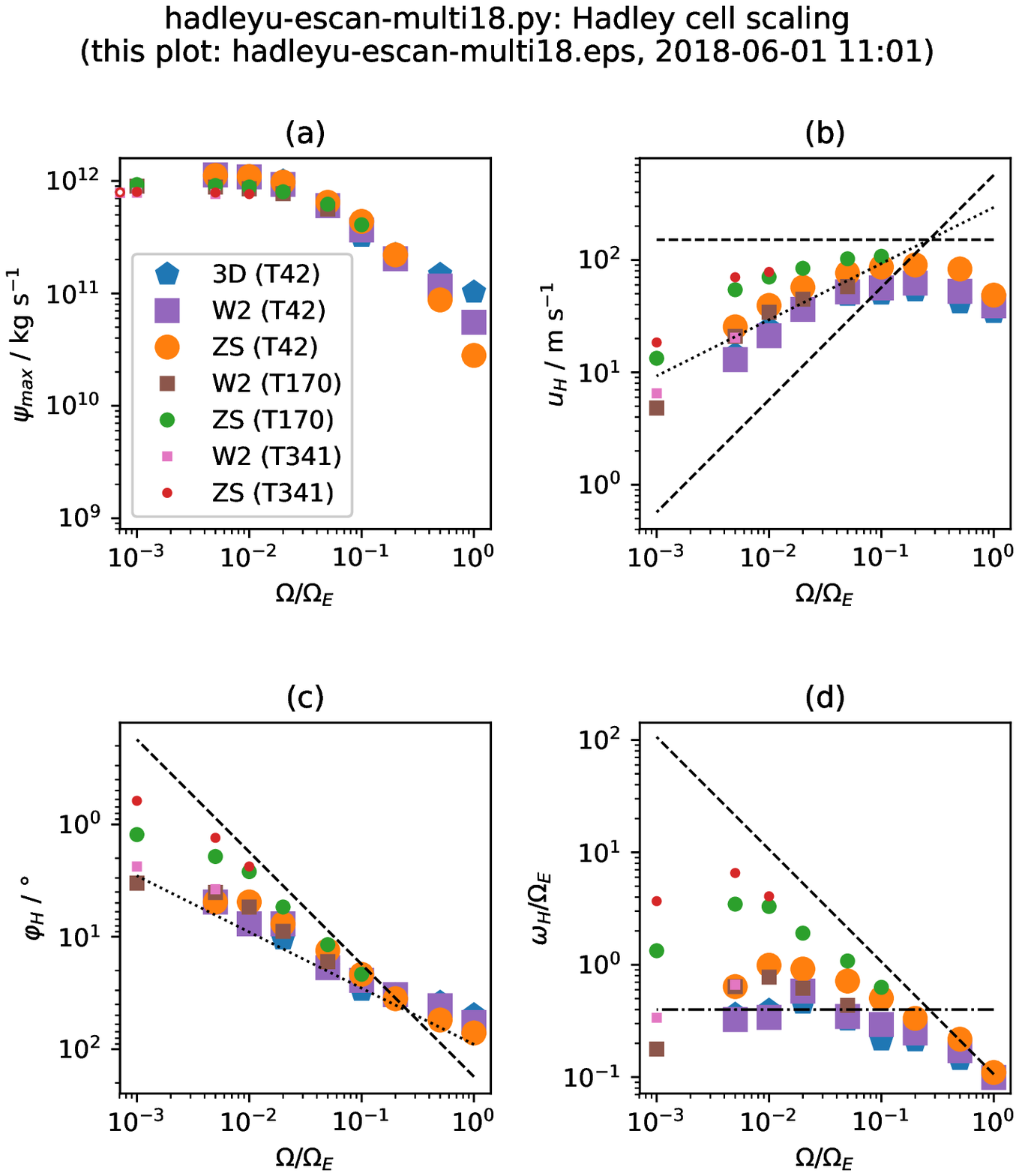} 
\caption{
(a) The magnitude of the hemispheric overturning circulation, as given by the maximum of the mass stream function $\psi$, versus planetary rotation rate $\Omega$;
(b) maximum (in latitude) of the zonal wind, $u_H$;
(c) Hadley cell boundary co-latitude, $\varphi_H$; (d) maximum zonal wind as an angular velocity around the planetary axis, $\omega_H\equiv u_H/a\sin\varphi_H$.
The boundary co-latitude is defined to be the co-latitude of the point of maximum zonal wind, indicated by the `+' markers in the previous two figures.
Various data sets are plotted, indicated in the legend of panel (a): fully 3D
simulations, at T42 resolution (pentagons); simulations truncated to zonal wavenumbers $m = \{0, 1, 2\}$ only, designated W2 (squares), at various resolutions (differentiated by marker size); and zonally symmetric simulations truncated to zonal wavenumber $m = 0$ only, designated ZS (discs), again at various resolutions.
In panel (a) the limiting value from W2 and ZS simulations at exactly $\Omega=0$ is shown (as hollow points on the vertical axis). In all cases the data are first processed by taking the 350-day zonal mean.
The dashed lines show the low-$\Omega$ limit of the HH theory; in panel (b) there are two such lines corresponding to the discontinuity in $u$: the upper line is on the equatorward side and the lower line is on the poleward side.
The dotted lines show the low-rotation limit of the modified theory with continuous $u$. In panel (d), the lower dashed line and the dotted line coincide, and are shown dot-dashed.}
\label{scalings}
\end{figure}

\subsection{Testing the theoretical scalings}
\label{testing}

We now compare the theoretical scalings for the HH theory and the continuous-$u$ variant with our numerical results; the comparison is shown in Fig.\ \ref{scalings}. The data points plotted include many more runs than those shown in Figs.\ \ref{uplot} and \ref{mplot}. The profiles plotted in Fig.\ \ref{hprofiles} also include more runs than Figs. \ref{uplot} and \ref{mplot}, and use the same color code as Fig. \ref{scalings}. The three rows of Figs. \ref{uplot} and \ref{mplot} correspond to a subset of rotation rates from the three T42 data sets in Fig.\ \ref{scalings} (indicated by the largest markers). Fig.\ \ref{scalings} includes intermediate and lower rotation rates, and also higher resolution runs (smaller markers) for the reduced (W2: $m \leq 2$ only) and zonally symmetric (ZS: $m=0$ only) data sets.

Our simulations use a hydrostatic primitive model rather than a Boussinesq model.  \cite{CPM} developed the Held--Hou theory for a compressible atmosphere with a more sophisticated radiative model, and found similar scaling behavior (but did not focus on low rotation rates). Here we take $\Delta_H gH\rightarrow \Delta T_y R_\mathrm{d}$, where $\Delta T_y = 60$ K is the Held--Suarez equator-pole temperature difference at the surface and $R_\mathrm{d} = 287$ J kg$^{-1}$ K$^{-1}$ is the gas constant for dry air. (Since the panels of Fig.\ \ref{scalings} are log-log plots, this value sets the intercepts only; it does not affect the slopes.)

The Held--Suarez forcing also features a stratospheric cap temperature of 200 K, below which the forcing does not drop at higher altitudes. We have not  needed to take additional account of this; it effectively sets the tropopause height at about the pressure scale height  (for Earth).
A general theory for other planets will need to pay more attention to this distinction; on Venus for example there are several scale heights between the tropopause and the surface.

Figs. \ref{uplot} and \ref{mplot} mark with a `+' sign in each panel the northern hemisphere maximum of $u$. In the zonally symmetric theories discussed above, the maximum $u$ occurs at the boundary between the low- and high-latitude regions, and so we take this maximum to define the boundary (co-)latitude in our simulations.

We tested these theories by performing a scan in $\Omega$ 
and plotting: the boundary co-latitude $\varphi_H$; the maximum zonal wind $u_H$; and
the maximum zonal wind expressed as an angular velocity around the planetary axis, $\omega_H\equiv u_H/a\sin\varphi_H$.
These are shown in panels (b)-(d) of Fig.\ \ref{scalings}, along with the predicted low-$\Omega$ scalings from section \ref{low-rot}, also shown as gray lines in the same panels of Fig. \ref{theory-scalings}.

In general we see that the simulations fall between the predictions of the two sets of scalings. In panel (c), the boundary co-latitude follows the HH scaling quite well in the low-rotation limit for the ZS simulations, provided sufficient resolution is used (higher T number). The 3D simulations appear closer to the continuous-$u$ prediction, but this result is not conclusive because of their limited resolution. 
Panels (b) and (d) tend to corroborate this story to some degree; in all simulations the maximum zonal wind does eventually decrease as rotation rate decreases, but it is sustained at a higher level, which is especially clear when viewed as an angular velocity in panel (d), for the zonally symmetric simulations at high resolution. 
We cannot be certain that the runs at the lowest nonzero rotation rate, $\Omega = \Omega_E/1000$, are converged in resolution even for the highest resolution shown here, so we include these points for completeness only.   It is interesting that at the next lowest rotation rate, $\Omega = \Omega_E/200$, the W2 runs do appear to be very well converged in resolution (pink square overlaying brown square), and much closer to the continuous-$u$ theory. It is a question for further work to determine if this is indeed representative of fully 3D runs that are converged in resolution.

\subsection{Further discussion}
\label{reconciling}

Fig. \ref{scalings}(a) shows the magnitude of the overturning circulation in the Hadley cell, $\psi_{\max}$, where $\psi$ is the mass stream function. On this panel are also shown, as (two coincident) hollow points on the vertical axis, the results from W2 and ZS simulations at exactly
$\Omega = 0$. 
These simulations have $u = 0$ everywhere, so they do not appear on the other three panels of Fig. \ref{scalings}.  Fig. \ref{scalings}(a) supports the point made earlier, that the overturning circulation is in good agreement between all three sets of runs (3D, W2 and ZS), with this agreement improved at low rotation rates and converging to a nonzero value at $\Omega = 0$.

\begin{figure}[t]
\centering
\includegraphics[trim={25 20 10 30},clip,scale=0.75]{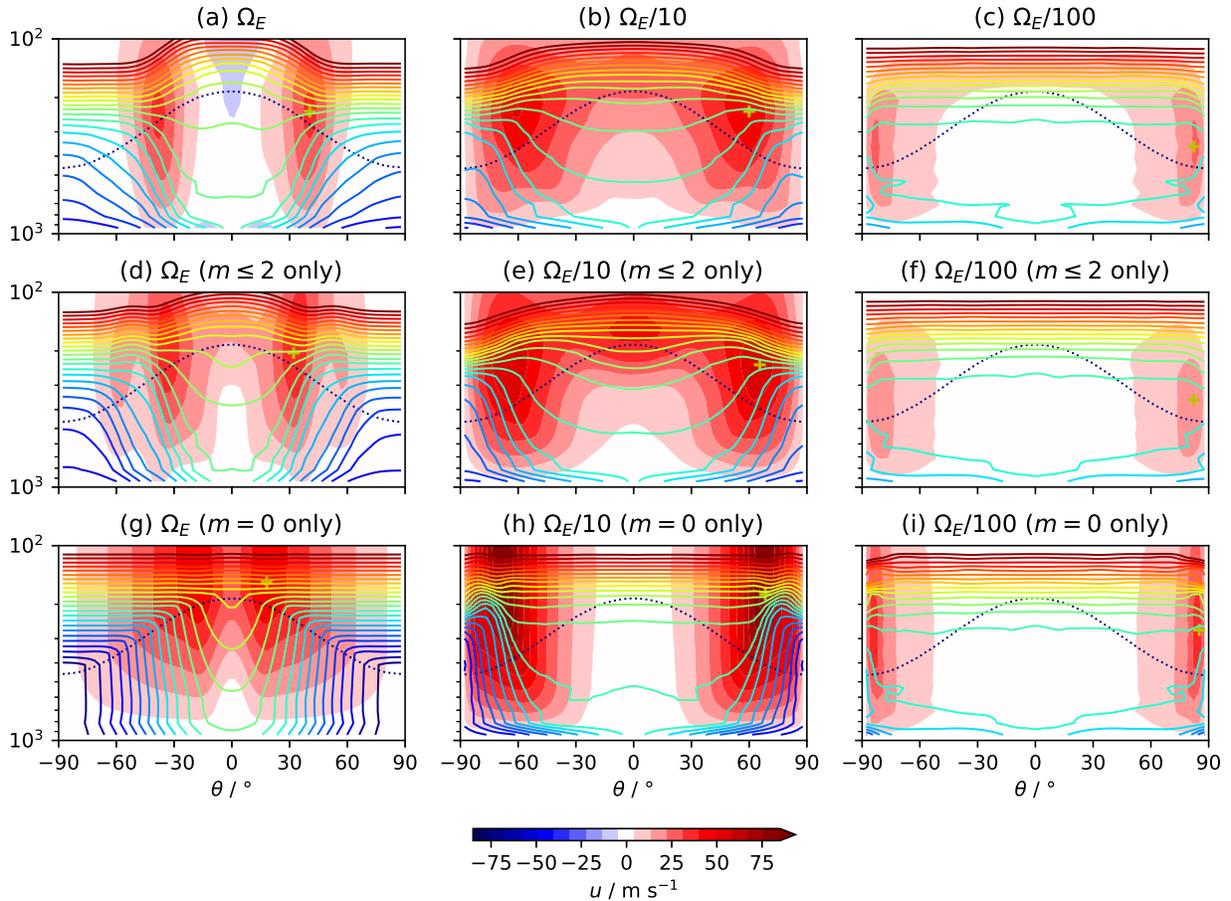} 
\caption{As Fig. \ref{uplot}, but showing potential temperature $\Theta$ instead of $T$ (solid contour lines, from 260 K to 375 K in 5 K intervals, matching Fig. 1(b) of \citealt{HS}).
}
\label{uplot_ptl}
\end{figure}

Estimates of the overturning time may be obtained from $m_H/\psi_{\max}$, where $m_H$ is the mass of the Hadley cell, or $a\theta_H/v$, or $H/w$. In the low-$\Omega$ limit, these are all of similar order to the Held--Suarez damping time scale (which is 40 days in the bulk of the atmosphere). Thus, in this limit the circulation is not in the very weak regime considered by HH; it is strong enough to drive the temperature towards the dry adiabatic lapse rate of constant potential temperature, as is demonstrated by Fig. \ref{uplot_ptl}.

\begin{figure}[t]
\centering
\includegraphics[trim={10 25 10 275},clip,scale=0.85]{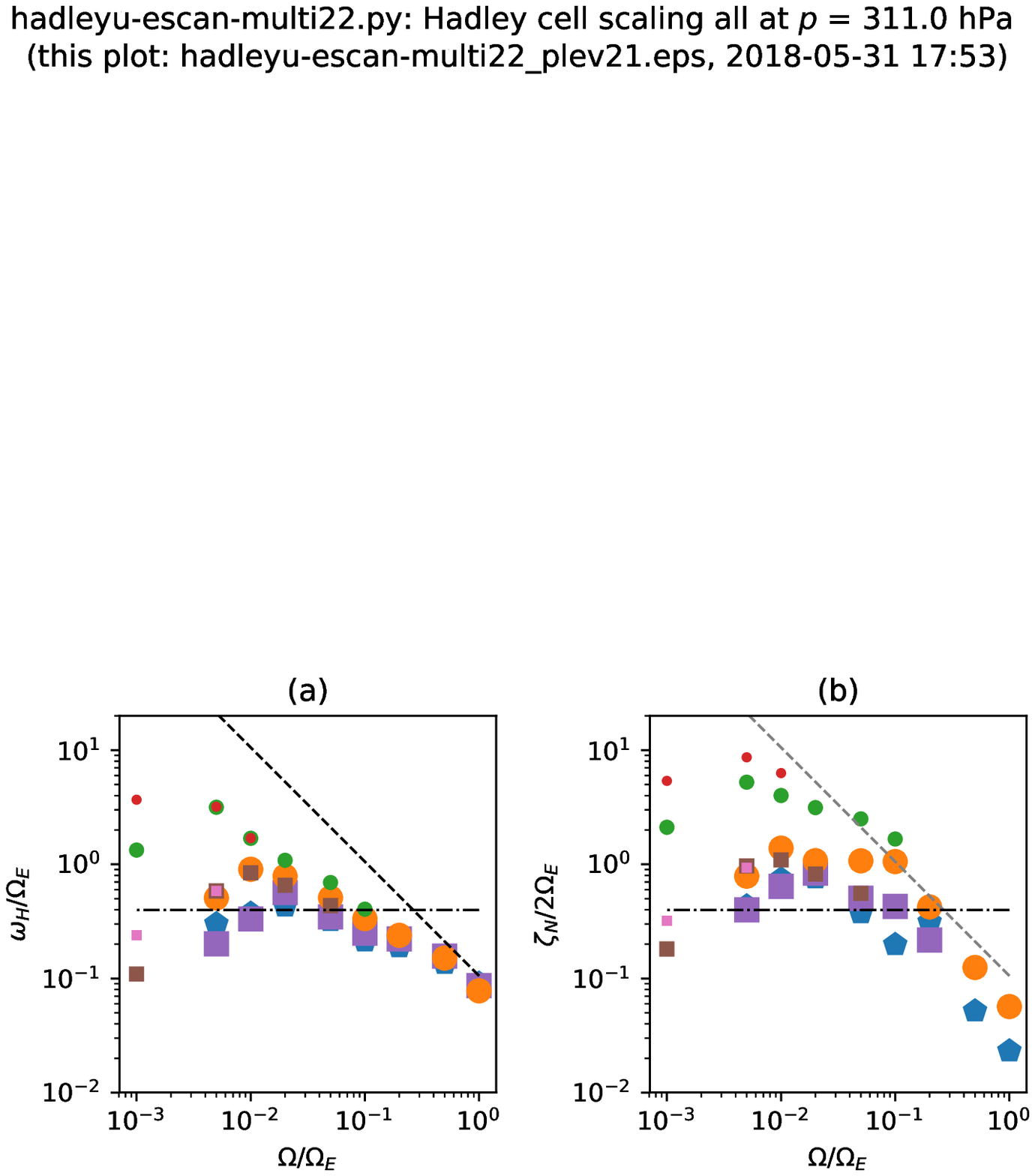} 
\caption{
As Fig. \ref{scalings}, but considering only the model pressure level $\sigma p_{\mathrm{ref}} = 311$ hPa.
(a) The maximum zonal wind as an angular velocity around the planetary axis, $\omega_H$;
(b) the relative vorticity at the north pole, $\zeta_N$, normalized to give an angular velocity in the same units of $\Omega_E$.
As in Fig. \ref{scalings}(d), the dashed line here represents the angular velocity in the HH theory on the equatorward side of the Hadley cell boundary; since the pole is on the other side, this line is shown in gray in panel (b).}
\label{scalings_plev21}
\end{figure}

We noted above that more general modifications to the polar temperature may be possible within the same theoretical framework, but would change the form of $u_E$ and hence the polar vorticity. 
In fact, whilst Fig. \ref{scalings}(d) clearly shows the boundary angular velocity to be sustained within an order of magnitude of $\Omega_E$, and therefore many orders of magnitude above $\Omega$ at the low-rotation end,  the profile plots in Fig. \ref{hprofiles} exhibit scant support for rigid-body rotation in the whole of the polar region, which is common to both of the zonally symmetric theories considered here (and indicated by the gray dot-dashed line in the middle row of Fig. \ref{hprofiles}). They do, however, suggest that the continuous-$u$ theory may have some validity. It is especially evident in 
column (b) of these profile plots that where the maximum $u$ occurs, the gradient of $u$ on either side is similar; and there is a distinct plateau in the angular velocity (middle row) and the vorticity (bottom row) at that point. One may surmise that in the zonally symmetric cases (yellow, green and red curves), the temperature that was raised to achieve continuity at the boundary is then required to return towards the equilibrium temperature at the pole, and is thus steepened, resulting in the even higher angular velocities and vorticities observed poleward. In the zonally symmetric case there can be no cross-polar flow. In the 3D and W2 cases, however, there can be, and so the temperature is not as constrained to steepen, and the poleward vorticity can remain close to the boundary value. 
The relative steepening of the temperature profile in the ZS case can be seen in the middle column of Fig. \ref{uplot}.
For all the runs, Fig. \ref{scalings_plev21}(a) shows the angular velocity at the point of maximum zonal wind taken over the single model level shown in Fig. \ref{hprofiles}, and Fig. \ref{scalings_plev21}(b) shows the vorticity evaluated at the north pole itself (in equivalent units). In the higher-resolution ZS cases (green and red discs), the polar vorticity is elevated close to the level corresponding to rigid-body rotation at the \emph{equatorward} boundary value from the HH theory (dashed line), whereas the other cases are closer to the predicted poleward value (dot-dashed line).

\begin{figure}[t]
\centering
\includegraphics[trim={25 20 10 150},clip,scale=0.75]{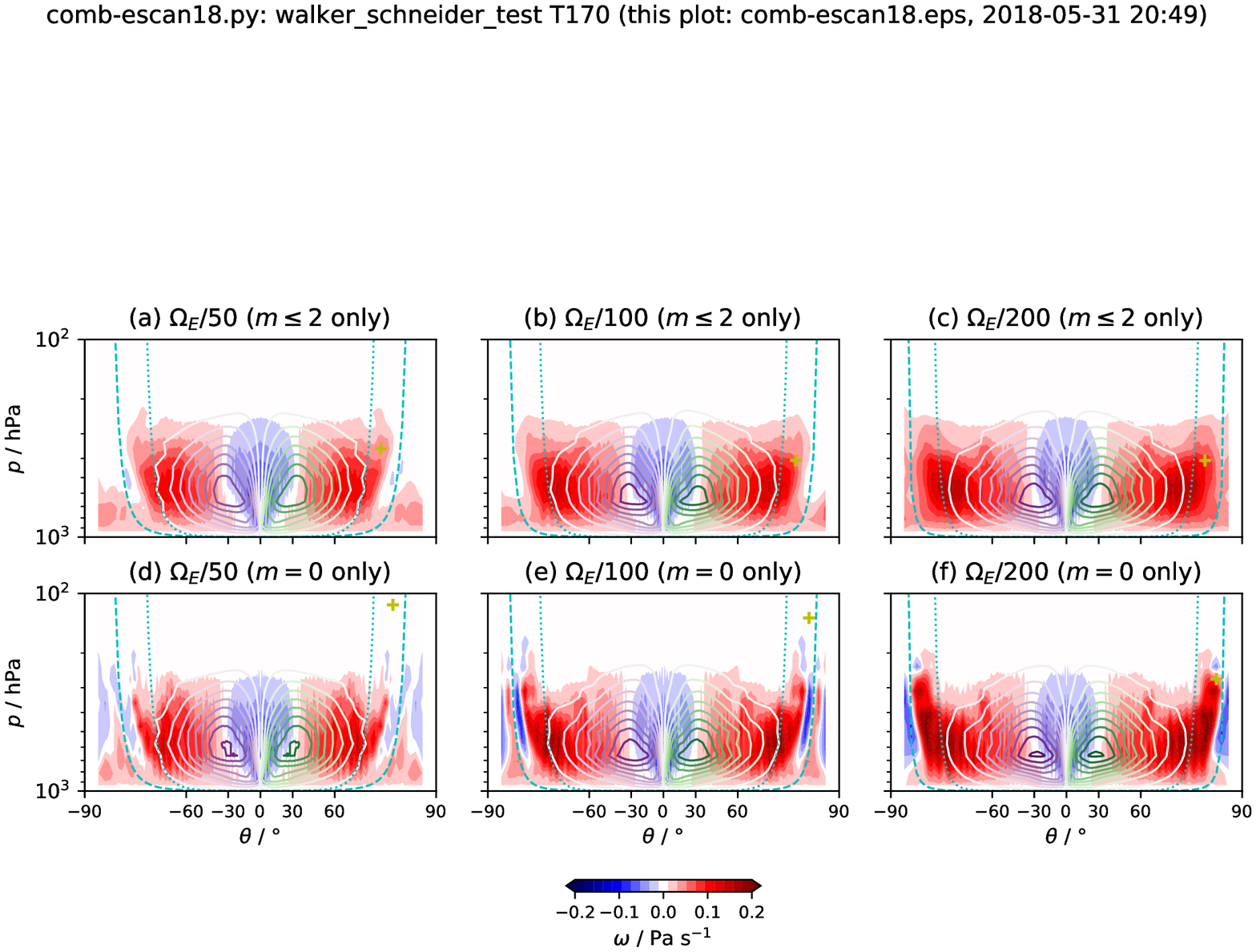} 
\caption{The vertical pressure velocity $\omega$ (shaded colors between contour levels) for a different range of planetary rotation rates $\Omega$ (labelled above each panel). As in Fig. \ref{mplot}, the yellow `+' sign marks the maximum $u$, and the stream function contour scale is also unchanged, but the expanded latitude scale of Fig. \ref{hprofiles} is used. These cases are all at T170 resolution, W2 (top row) and ZS (bottom row) only.
}
\label{wplot}
\end{figure}

We do not have a full understanding of the polar overturning circulation, but it may be characterized in somewhat general terms by considering the contour plots of the vertical pressure velocity $\omega\equiv Dp/Dt$ in Fig. \ref{wplot}. $\omega$ is shown in order to bring out the details near the pole, where the circulation contributes too little to the mass stream function to be visible in the stream function contours.
However, it is clear from the stream function contours in Fig. \ref{mplot} that there is a reversed overturning circulation at high latitudes in the 3D and W2 cases at $\Omega=\Omega_E/10$ (panels (b) and (e) of Fig. \ref{mplot}), and this is no longer evident at $\Omega=\Omega_E/100$ (panels (c) and (f) of Fig. \ref{mplot}). In Fig. \ref{wplot}, the upward vertical wind on the poleward side of such a reversed circulation would show as blue ($\omega < 0$). For the W2 cases, a trace of it can still just be seen at $\Omega=\Omega_E/50$ in Fig. \ref{wplot}(a), but not at $\Omega=\Omega_E/100$ in Fig. \ref{wplot}(b), and at $\Omega=\Omega_E/200$ in Fig. \ref{wplot}(c) the downward Hadley circulation ($\omega> 0$, red) extends all the way poleward. However, for the ZS cases (bottom row of Fig. \ref{wplot}), a reversed poleward circulation persists, and further structure is evident especially in panel (e). 
Note that since all of the features discussed here occur only within the troposphere (the upper part of all the panels is white), they are distinguished from the `deep' stratosphere-encompassing circulation observed by \cite{CPM}.

Neither zonally symmetric theory could apply exactly to a full GCM simulation, in which one would not expect actual discontinuities to be maintained. Since eddies may in general play a role in removing discontinuities (or opposing their creation), via turbulent diffusion for example, but such a mechanism is not possible in a zonally symmetric simulation, some other mechanism must operate in this case---and may or may not also be the dominant mechanism operating in 3D simulations. Here, we see that the ZS simulations do seem to be distinguished from the other two sets, but it is also  interesting that the apparently under-resolved ZS simulations at T42 are in better agreement with the 3D and W2 cases; so perhaps grid-scale numerical dissipation is mimicking the role of small-scale eddies.

As a final note, we observe that the 3D cases exhibit both $u(0) < 0$ and $u(0) > 0$ on occasion (blue curves for $\Omega=\Omega_E$ and $\Omega=\Omega_E/10$ respectively in Fig. \ref{hprofiles}).

\section{Implications for modeling Venus}
\label{venus}

Although we do not study Venus specifically in this paper, we wish to point out a couple of relevant implications. First, in these simulations (which are all at Earth's surface pressure---that of Venus is two orders of magnitude higher), by the time $\Omega$ has
decreased to $\Omega_E/100$, which is still a faster rotation rate than Venus, the maximum zonal wind is already considerably lower than for Earth, even for the 3D simulations, whereas in reality Venus is observed to have faster zonal winds than  Earth (e.g. in Venus Express data analyzed by \citealt{Sanchez}). We find the same effect with higher surface pressures too (not included here). Some other mechanism is required in order to sustain the zonal winds on Venus,
for example diurnal and semi-diurnal thermal tides (e.g. \citealt{FL74}), or effects arising from a better treatment of radiative forcing.

Our results may, however, contain clues to an explanation of Venus's strong polar vortex, and of its `cold collar' (e.g. \citealt{Ando}), which is a minimum in temperature (at a given altitude) at a latitude below the pole. 
In Fig. \ref{uplot}, at the highest rotation rates (left column) the temperature decreases poleward, as does the forcing temperature, whereas at the lowest rotation rates (right column) it is flattened across most of the planet and increases poleward aloft at very high latitude. Although our present results contain no strong examples of temperature first decreasing and then increasing poleward in the same case, it seems plausible that that may occur due partly to the type of transition seen here.

\section{Conclusions}
\label{conclusions}

In this paper we have studied the zonally symmetric theory of \citet{HH} and \citet{Hou84}, and a variant with continuous zonal wind as its boundary matching condition, focusing mainly on the limit of low planetary rotation rate. We find that the original theory and this variant have different scalings for the boundary co-latitude and boundary zonal wind, as well as the different continuity properties. Simulations using a GCM are found to have smoother temperature profiles than either theory, and also feature an overturning circulation poleward of the point of maximum zonal wind, which is absent in both theories. It is the presence of this overturning 
circulation in the polar region (in both 3D and zonally symmetric simulations) that allows the zonal velocity profile too to be smoother than the original theory, and that removes the temperature discontinuities of the variant theory, without the need for viscous or diffusive smoothing. 
Resolved zonally symmetric simulations fall between the two sets of theoretical scalings, and have a faster polar zonal flow than both theories, consistent with a steepened polar temperature profile. The 3D simulations fall closer to the predictions of the variant theory and the theoretical polar zonal flow, perhaps because allowed cross-polar flow lifts the requirement to steepen the polar temperature profile to the same extent. Even in the zonally symmetric simulations, the maximum $u$ falls with falling $\Omega$ at low rotation rates. This is easier to reconcile with the $u = 0$, $\Omega = 0$ case than the low-$\Omega$ limit of the original theory (which retains nonzero $u$), and indicates that an additional mechanism is required to sustain strong zonal winds for very low planetary rotation rates. The overturning circulation in the Hadley cell increases to a finite maximum as $\Omega\rightarrow 0$, and is insensitive to zonal symmetry. The maximum overturning circulation is strong, in the sense that it drives the temperature profile close to a state of constant potential temperature.

More generally, our results show that the zonally symmetric theory is able to predict the qualitative behavior of the Hadley cell, and the associated zonal wind, as planetary rotation is varied over quite a wide range. Both the original theory and its variant have quantitative shortcomings.   Zonally-symmetric numerical simulations (which is what the theories are constructed to describe) do not show a discontinuity in either zonal wind or temperature, at either low or high rotation rates, nor even a particularly fast variation of either quantity that might have been indicative of a discontinuity smoothed away by viscosity or diffusion. The reason for the smoothness seems to be that the region poleward of the edge of the Hadley cell is not in radiative equilibrium; rather, it has a non-zero circulation that enables the temperature to blend continuously with that in the Hadley cell region. 

The three-dimensional simulations further differ from the theoretical predictions in two main ways. First, at high rotation rates, baroclinic eddies extract momentum (and heat) from the low-latitude Hadley cell and the zonal wind does not quantitatively follow the angular-momentum-conserving profile, as is well known. Second, at low rotation rates the three-dimensional simulations (even with the zonal structure limited to wavenumbers 1 and 2) produce superrotation.  Superrotation aside,  the zonally symmetric theory is a better model of the three-dimensional circulation at low rotation rates than at high,  because the lack of baroclinic eddies allows  angular momentum to be better conserved.  However, the reduction of the zonal wind at very low rotation rates is neither exactly predicted nor completely understood.

\vspace{5mm}
\acknowledgments

This work has been supported by the Leverhulme Trust. 
We appreciate productive discussions with R. Geen, M. Jucker, F. Poulin,  P. Maher,  A. Paterson,  J. Penn and S. Thomson. The SHTns library \citep{SHTns} was used to obtain the vorticity at the pole from its spectral representation.

\appendix

\appendixtitle{Choice of $H$ for comparison with numerical simulations}

In subsection \ref{numerical}\ref{testing} we make the correspondence $\Delta_H gH\rightarrow \Delta T_y R_\mathrm{d}$ to go from the theory to the numerical model. In this Appendix we obtain a more precise correspondence. We start from the following equation for thermal wind gradient balance in the numerical model:
\begin{equation}
\frac{\partial}{\partial\ln p}\left(fu + \frac{u^2\tan\theta}{a}\right) = \frac{R_d}{a}\frac{\partial T}{\partial\theta}.
\end{equation}
Integrating from the surface, where $p=p_s$ and $u=0$, gives at pressure level $p$,
\begin{equation}
fu + \frac{u^2\tan\theta}{a} = \frac{R_d}{a}\frac{\partial}{\partial\theta}\int^p_{p_s} T d(\ln p).
\end{equation}
To obtain the equilibrium zonal wind (equivalent to $u_E$ in the main paper) we evaluate the right-hand side using $T = T_{\mathrm{eq}}$, the Held--Suarez forcing \citep{HS}. Ignoring both the stratospheric cap temperature and the small term in $\ln(p/p_0)$, this is
\begin{equation}
T_{\mathrm{eq}}=[315\,\mathrm{K} - \Delta T_y\sin^2\theta]\left(\frac{p}{p_0}\right)^\kappa,
\label{simpleHS}
\end{equation}
where $p_0 = 1000$ hPa and $\kappa=2/7$. With $p_s=p_0$, we then obtain the same solution as \eqref{uE}, with the correspondence
\begin{equation}
\Delta_H gH\rightarrow \frac{1}{\kappa}\left[1-\left(\frac{p}{p_0}\right)^\kappa\right]\Delta T_y R_\mathrm{d}.
\label{mapping}
\end{equation}
Conveniently, the pressure at which this extra prefactor $(1/\kappa)[1-({p}/{p_0})^\kappa]$ is equal to unity, $p/p_0 = (5/7)^{7/2} = 0.308$, is
a good choice for a representative tropopause pressure (the actual level of maximum $u$ differs between runs, as indicated by the yellow `+' signs in the figures), and the closest model level to this has been selected for Figs. \ref{hprofiles} and \ref{scalings_plev21}. Justifying the simplifications leading to \eqref{simpleHS}, there is good agreement between the orange curve and the dot-dashed curve in the top-left panel of Fig. \ref{hprofiles}, as expected from the correspondence between $T$ and $T_{\mathrm{eq}}$ in Fig. \ref{uplot}(g).

The cyan curves in Figs. \ref{mplot} and \ref{wplot} are generated using \eqref{R_HH} or \eqref{newcrossover} as appropriate, with $R$ dependent on $H$ according to \eqref{mapping}. The fact
that the boundary of the angular-momentum-conserving region (in Fig. \ref{mplot}) approximately follows these curves, means that the particular choice of model level selected for comparison is not too critical.

\bibliographystyle{ametsoc2014}
\bibliography{references}

\begin{thebibliography}{21}
\providecommand{\natexlab}[1]{#1}
\providecommand{\url}[1]{\texttt{#1}}
\renewcommand{\UrlFont}{\rmfamily}
\providecommand{\urlprefix}{URL }
\expandafter\ifx\csname urlstyle\endcsname\relax
  \providecommand{\doi}[1]{doi:\discretionary{}{}{}#1}\else
  \providecommand{\doi}{doi:\discretionary{}{}{}\begingroup
  \urlstyle{rm}\Url}\fi
\providecommand{\eprint}[2][]{\url{#2}}

\bibitem[{Ando et~al.(2016)Ando, Sugimoto, Takagi, Kashimura, Imamura,, and
  Matsuda}]{Ando}
Ando, H., N.~Sugimoto, M.~Takagi, H.~Kashimura, T.~Imamura, and Y.~Matsuda,
  2016: The puzzling {Venusian} polar atmospheric structure reproduced by a
  general circulation model. \textit{Nature Communications}, \textbf{7}, 10398.

\bibitem[{Caballero et~al.(2008)Caballero, Pierrehumbert,, and Mitchell}]{CPM}
Caballero, R., R.~T. Pierrehumbert, and J.~L. Mitchell, 2008: Axisymmetric,
  nearly inviscid circulations in non-condensing radiative-convective
  atmospheres. \textit{Quart.\ J.\ Roy.\ Meteor.\ Soc.}, \textbf{134},
  1269--1285.

\bibitem[{Covey et~al.(1986)Covey, Pitcher,, and Brown}]{Covey}
Covey, C., E.~J. Pitcher, and J.~P. Brown, 1986: General circulation model
  simulations of superrotation in slowly rotating atmospheres: Implications for
  {Venus}. \textit{Icarus}, \textbf{66}, 380--396.

\bibitem[{Fels and Lindzen(1974)Fels, and Lindzen}]{FL74}
Fels, S.~B., and R.~S. Lindzen, 1974: The interaction of thermally excited
  gravity waves with mean flows. \textit{Geophysical Fluid Dynamics},
  \textbf{6}, 149--191.

\bibitem[{Ferrel(1859)}]{Ferrel59}
Ferrel, W., 1859: The motion of fluids and solids relative to the {Earth's}
  surface. \textit{Math.\ Monthly}, \textbf{1}, 140--148, 210--216, 300--307,
  366--373, 397--406.

\bibitem[{Geen et~al.(2018)Geen, Lambert,, and Vallis}]{Geen}
Geen, R., F.~H. Lambert, and G.~K. Vallis, 2018: Regime change behavior during
  {Asian} monsoon onset. \textit{J.\ Climate}, \textbf{31}, 3327--3348.

\bibitem[{Hadley(1735)}]{Hadley35}
Hadley, G., 1735: Concerning the cause of the general trade-winds.
  \textit{Phil.\ Trans.\ Roy.\ Soc.}, \textbf{29}, 58--62.

\bibitem[{Held and Hou(1980)Held, and Hou}]{HH}
Held, I.~M., and A.~Y. Hou, 1980: Nonlinear axially symmetric circulations in a
  nearly inviscid atmosphere. \textit{J.\ Atmos.\ Sci.}, \textbf{37}, 515--533.

\bibitem[{Held and Suarez(1994)Held, and Suarez}]{HS}
Held, I.~M., and M.~J. Suarez, 1994: A proposal for the intercomparison of the
  dynamical cores of atmospheric general circulation models. \textit{Bull.\
  Amer.\ Meteor.\ Soc.}, \textbf{75}, 1825--1830.

\bibitem[{Hide(1969)}]{Hide69}
Hide, 1969: Dynamics of the atmospheres of the major planets with an appendix
  on the viscous boundary layer at the rigid bounding surface of an
  electrically-conducting rotating fluid in the presence of a magnetic field.
  \textit{J.\ Atmos.\ Sci.}, \textbf{26}, 841--853.

\bibitem[{Hou(1984)}]{Hou84}
Hou, A.~Y., 1984: Axisymmetric circulations forced by heat and momentum
  sources: A simple model applicable to the {Venus} atmosphere. \textit{J.\
  Atmos.\ Sci.}, \textbf{41}, 3437--3455.

\bibitem[{Lindzen and Hou(1988)Lindzen, and Hou}]{LindzenHou}
Lindzen, R.~S., and A.~Y. Hou, 1988: Hadley circulations for zonally averaged
  heating centered off the equator. \textit{J.\ Atmos.\ Sci.}, \textbf{45},
  2416--2427.

\bibitem[{Lorenz(1967)}]{Lorenz67}
Lorenz, E.~N., 1967: \textit{The Nature and the Theory of the General
  Circulation of the Atmosphere}, {WMO} Publications, Vol. 218. World
  Meteorological Organization.

\bibitem[{Mitchell and Vallis(2010)Mitchell, and Vallis}]{MV}
Mitchell, J.~L., and G.~K. Vallis, 2010: The transition to superrotation in
  terrestrial atmospheres. \textit{J.\ Geophys.\ Res.}, \textbf{115}, E12008.

\bibitem[{Plumb and Hou(1992)Plumb, and Hou}]{PlumbHou}
Plumb, R.~A., and A.~Y. Hou, 1992: The response of a zonally symmetric
  atmosphere to subtropical thermal forcing: Threshold behavior. \textit{J.\
  Atmos.\ Sci.}, \textbf{49}, 1790--1799.

\bibitem[{S\'anchez-Lavega et~al.(2008)}]{Sanchez}
S\'anchez-Lavega, A., et al., 2008: Variable winds on {Venus} mapped in three
  dimensions. \textit{Geophys.\ Res.\ Lett.}, \textbf{35}, L13204.

\bibitem[{Schaeffer(2013)}]{SHTns}
Schaeffer, N., 2013: Efficient spherical harmonic transforms aimed at
  pseudospectral numerical simulations. \textit{Geochemistry, Geophysics,
  Geosystems}, \textbf{14}, 751--758.

\bibitem[{Schneider(1977)}]{ESchneider77}
Schneider, E.~K., 1977: Axially symmetric steady-state models of the basic
  state for instability and climate studies. part ii. nonlinear calculations.
  \textit{J.\ Atmos.\ Sci.}, \textbf{34}, 280--296.

\bibitem[{Thomson(1892)}]{Thomson1892}
Thomson, J., 1892: Bakerian lecture.---{On} the grand currents of atmospheric
  circulation. \textit{Phil.\ Trans.\ Roy.\ Soc.\ Lond.\ A}, \textbf{183},
  653--684.

\bibitem[{Vallis(2017)}]{AOFD2}
Vallis, G.~K., 2017: \textit{Atmospheric and Oceanic Fluid Dynamics, 2nd ed.}
  Cambridge University Press.

\bibitem[{Vallis et~al.(2018)}]{Isca}
Vallis, G.~K., et al., 2018: Isca, v1.0: a framework for the global modelling
  of the atmospheres of {Earth} and other planets at varying levels of
  complexity. \textit{Geosci. Model Dev.}, \textbf{11}, 843--859.

\end{thebibliography}

\newpage

\end{document}